# On Gröbner-Shirshov bases for Markov semirings

*Xiaohui NIU*[1] *Wenxi LI*[†2] *Zhongzhi WANG*[3]

( School of Microelectronics and Data Science, AnHui University of Technology, Ma'anshan, 243002, China)

**Abstract:** In order to investigate the relationship between Shannon information measure of random variables, scholars such as Yeung utilized information diagrams to explore the structured representation of information measures, establishing correspondences with sets. However, this method has limitations when studying information measures of five or more random variables.

In this paper, we consider employing algebraic methods to study the relationship of information measures of random variables. By introducing a semiring generated by random variables, we establish correspondences between sets and elements of the semiring. Utilizing the Gröbner-Shirshov basis, we present the structure of the semiring and its standard form. Furthermore, we delve into the structure of the semiring generated under Markov chain conditions (referred to as Markov semiring), obtaining its standard form.

## 1. Introduction

In order to find simplified solutions to difficult problems in information theory, Hu[5] initially studied the set-theoretic structure of Shannon information measures. It was shown in his paper that every information identity implies a set identity via a substitution of symbols, and the corresponding relationship between set theory and Shannon information measures was also established. Throughout many years, the use of Venn diagrams to represent the structure of Shannon information measures for two or three random variables has been suggested by lots of authors, for instances, Reza [9], Abramson [1]and Papoulis [8]. Meanwhile, in 1991, the I-measure introduced by Yeung [12] is another way to explain it. Therefore, the rich computing tools and methods in set theory can be applied to information theory. By using this method, Yeung visualizes the relationship between Shannon information measures of four or less random variables with an information diagram.

For information diagram, there are also applications in computer science and technology, physics, mechanics and engineering. In 2003, by using information diagram, Fry [3] presented a brief overview of the design process of systems theory and provided a quantitative basis for neural computation. In 2010, Valverde-Albacete et al. [11] gave the balance equation by studying the relationship between information measures, and the balance equation suggests an information diagram somewhat more complete than what is normally used for the relations between the entropies of two variables as depicted in [13]. In 2011, James et al. [7] used information diagrams to deepen their understanding of multivariate information defined in the multivariate information measure, and analyzed the information embedded in discrete-valued random time series. In 2016, Rosas et al. [10] analyzed the interaction between three or more random variables and introduced an axiomatic framework to decompose the joint entropy (The framework describes share information). In 2017, Ince [6] derived the decomposition of multivariate entropy through the analysis of information measures, and the resulting decomposition provides useful tools for practical data analysis. However, there is also certain limitation in Venn diagram. When the number of random variables is more than five, using the information diagrams to show the relationship between their Shannon information measures becomes difficult.

[1]E-mail address:nxiaohui0219@163.com

[2]†Corresponding author's E-mail address: wxli@ahut.edu.cn.

[3]E-mail address: zhongzhiw@126.com.

In this paper, for the algebraic structure of sets with intersection and union operations, we introduce semirings generated by random variables. In 2013, Bokut et al. [2] derived Gröbner-Shirshov bases of semirings and commutative semirings, and the normal forms of semirings can be obtained by using the C-D lemma. For free semirings generated by $n$ random variables, we use Gröbner-Shirshov bases and Shirshov algorithm to give the algebraic expression of the semirings, and obtain their normal forms according to the C-D lemma.

Furthermore, in order to study the Markov chain, we introduce the Markov semiring and give the algebraic expression of the Markov semiring generated by five random variables and its normal form, in doing so, a general method for solving the Shannon information measures expression for $n$ random variables is obtained.

The rest of this article is organised as follows. In section 2, we give some concepts and properties of semiring and semiring algebra; In section 3, we give the algebraic structure of the free idempotent complement semirings generated by random variables and their normal forms; In section 4, we give the algebraic structure of the Markov semirings and their normal forms.

Throughout this paper, we follow [2] and denote $Rig[X]$ the commutative semiring generate by a set $X$, $\mathbb{N}$ the set of natural numbers, $Rig[X|R]$ the commutative semiring with generators $X$ and define realtions $R$ i.e. $Rig[X|R] = Rig[X]/\equiv_{\rho(R)}$, where $\equiv_{\rho(R)}$ is the congruence of $Rig[X]$ generated by $R$. The free semirings, semiring algebras and its related concepts (ideals, monomial orders, least common multiplier, normal form, monic polynomials) are all derived from [2]. General background material can be found in[2], [4] and[13].

## 2. Semiring and semiring algebra

Noting that sets with intersection and union operations can form semirings, we first give some concepts and properties of semiring algebras.

**Definition 2.1.** A semiring is a nonempty set $S$ on which operations of "$\circ$" and multiplication " $\cdot$ " have been defined such that the following conditions are satisfied:

1) $(S, \circ)$ is a commutative monoid with identity element $\theta$.
2) $(S, \cdot)$ is a monoid with identity element 1.
3) "$\cdot$" is distributive relative to $\circ$ form left and right.
4) $\theta \cdot s = s \cdot \theta = \theta$, for all $s \in S$.
5) $1_S \neq \theta$.

If $(S, \cdot)$ is commutative, then $S$ is called a commutative semiring.

**Definition 2.2.** [2, Definition 4.1] Let $<$ be a monomial ordering on $Rig[X]$. Let $f, g$ be two monic polynomials in $kRig[X]$ and $\overline{f} = u_1 \circ u_2 \circ \cdots \circ u_n$, $\overline{g} = v_1 \circ v_2 \circ \cdots \circ v_m$, where each $u_i, v_j \in [X]$. For any pair $(a, b) \in \{(a_{ij}, b_{ij}) \mid 1 \leqslant i \leqslant n, 1 \leqslant j \leqslant m$, where $a_{ij}, b_{ij} \in [X]\}$ such that $lcm.(u_i, v_j) = a_{ij}u_i = b_{ij}v_j$, we call $(f, g)_w = fa \circ u - bg \circ v$ the composition of $f$ and $g$ with respect to $w$, where $w = lcm_\circ(a\overline{f}, b\overline{g}) = a\overline{f} \circ u = b\overline{g} \circ v$.

In the above definition, $w$ is called the ambiguity of the composition. Obviously,

$$(f, g)_w \in Id(f, g) \quad and \quad \overline{(f, g)_w} < w,$$

where $Id(f, g)$ is the ideal of $kRig[X]$ generated by the set $\{f, g\}$.

**Definition 2.3.** [2, Definition 4.2] Suppose that $w$ is a monomial in $Rig[X]$, $R$ a set of monic polynomials in $kRig[X]$ and $h$ a polynomial. Then $h$ is trivial modulo $(R, w)$, and denoted by $h \equiv 0 \bmod (R, w)$, if $h = \sum_i \alpha_i a_i s_i \circ u_i$, where each $\alpha_i \in k$, $a_i \in [X]$, $u_i \in Rig[X]$ ,$r_i \in R$ and $a_i\overline{r_i} \circ u_i < w$.

The set $R$ is called a *Gröbner − Shirshov* basis in $kRig[X]$ if any composition in $R$ is trivial modulo $R$ and corresponding $w$.

Denote

$$Irr(R) = \{w \in Rig[X] | w \neq a\bar{r}b \circ u \ for\ any\ a, b \in [X], u \in Rig[X], r \in R\}.$$

A Gröbner-Shirshov basis $R$ in $kRig[X]$ is reduced if for any $r \in R$, $supp(r) \subseteq Irr(R \backslash r)$, where $supp(r) = u_1, u_2, \cdots, u_n$ if $r = \sum_{i=1}^{n} \alpha_i u_i$, $0 \neq \alpha_i \in k$, $u_i \in Rig[X]$.

If the set $R$ is a Gröbner-Shirshov basis in $kRig[X]$, then we call also $R$ is a Gröbner-Shirshov basis for the ideal $Id(R)$ or the algebra $kRig[X|R] := kRig[X]/Id(R)$.

Let $I$ be an ideal of $kRig[X]$. Then there exists a uniquely reduce Gröbner-Shirshov basis $R$ for $I$ [2, Theorem 3.5].

**Lemma 2.1.** [2, Theorem 4.4] *(C-D lemma for commutative semirings).*

*Let $R$ be a set of monic polynomials in $kRig[X]$ and $<$ a monomial ordering on $Rig[X]$. Then the following statements are equivalent:*

*(1) $R$ is a Gröbner-Shirshov basis in $kRig[X]$.*

*(2) $f \in Id(R) \Rightarrow \overline{f} = a\bar{r} \circ u$ for some $a \in [X], u \in Rig[X]$ and $r \in R$.*

*(3) Irr(R)= $\{w \in Rig[X] | w \neq a\bar{r} \circ u$ for any $a \in [X]$,$u \in Rig[X]$, $r \in R\}$ is a k-linear basis of $kRig[X|R] = kRig[X]/Id(R)$.*

## 3. **Idempotent complement semiring**

Let $X=\{x_1,\cdots,x_n\}, X^c=\{x_1^c,\cdots,x_n^c\}$ and $\tilde{X}=X\cup X^c$. Let $Rig[\tilde{X}]$ be a free commutative semiring generated by $\tilde{X}$. We define congruence relation $\equiv_\rho$ in $Rig[\tilde{X}]$ generated by $\{(x_i\cdot x_i^c,\theta),(x_i\circ x_i^c,1),(x_i\circ x_i,x_i),(x_i^c\circ x_i^c,x_i^c)\}$. In this section, we mainly study $Rig[\tilde{X}]/\equiv_\rho$.

Denote $Y=\{y_0,\cdots,y_{m-1}\}(m=2^n,n\in\mathbb{N})$, $k=\sum_{l=1}^{n} j_l 2^{l-1}$, $j_l\in\{0,1\}$ and $y_k=x_1^{k_1}\cdot\cdots\cdot x_n^{k_n}\in Rig[\tilde{X}]$, where

$$x_i^k=\begin{cases} x_i & k=1\\ x_i^c & k=0\end{cases} \tag{3.1}$$

Let $\equiv_{\rho_1}$ be a congruence relation in $Rig[\tilde{X}\cup Y]$ generated by $\{(x_i\cdot x_i^c,\theta),(x_i\circ x_i^c,1),(x_i\circ x_i,x_i),(x_i^c\circ x_i^c,x_i^c),(y_k,x_1^{k_1}\cdot\cdots\cdot x_n^{k_n})\}$. Since $Rig[\tilde{X}]/\equiv_\rho$ is also generated by $\tilde{X}\cup Y$, it can also be represented as $Rig[\tilde{X}\cup Y]/\equiv_{\rho_1}$.

Next, suppose that a congruence relation $\equiv_{\rho_2}$ is generated by $\{(y_k\cdot y_k,y_k),(y_j\cdot y_k,\theta),(y_k\circ y_k,y_k),(x_i,\sum^\circ_{j\in A_i} y_i),(x_i^c,\sum^\circ_{j\in A_i^c} y_j),(\sum^\circ_{j\in D} y_j,1),(1\circ y_k,1),(1\circ 1,1)\}$, where $\sum^\circ$ is represented as the sum of " $\circ$ " operation, $D=\{0,1,2,\cdots,m-1\}$,

$$A_i=\{\sum_{j=1}^{n} k_j 2^{j-1} | k_i=1,k_1,\cdots,k_{i-1},k_{i+1},\cdots,k_n\in\{0,1\}\}, \tag{3.2}$$

$$A_i^c=\{\sum_{j=1}^{n} k_j 2^{j-1} | k_i=0,k_1,\cdots,k_{i-1},k_{i+1},\cdots,k_n\in\{0,1\}\}. \tag{3.3}$$

**Proposition 3.1.** $\equiv_{\rho_1}=\equiv_{\rho_2}$

*Proof.* We first prove $\equiv_{\rho_2}\subseteq\equiv_{\rho_1}$. The proof falls naturally into 8 parts.

(1). Noticing that

$$\begin{aligned} y_k\cdot y_k &\equiv_{\rho_1}(x_1^{k_1}\cdot\cdots\cdot x_n^{k_n})\cdot(x_1^{k_1}\cdot\cdots\cdot x_n^{k_n})\\ &\equiv_{\rho_1}(x_1\cdot x_1)^{k_1}\cdot\cdots\cdot(x_n\cdot x_n)^{k_n}. \end{aligned}$$

If $x_i^c\circ x_i\equiv_{\rho_1}1$, we have $(x_i^c\cdot x_i)\circ(x_i\cdot x_i)\equiv_{\rho_1}x_i$ and $(x_i^c\cdot x_i^c)\circ(x_i\cdot x_i^c)\equiv_{\rho_1}x_i$. If $x_i^c\cdot x_i\equiv_{\rho_1}\theta$, we have $x_i\cdot x_i\equiv_{\rho_1}x_i$ and $x_i^c\cdot x_i^c\equiv_{\rho_1}x_i^c$. So, $y_k\cdot y_k\equiv_{\rho_1}y_k$.

(2).

$$\begin{aligned} y_k\cdot y_j(k\neq j) &\equiv_{\rho_1}(x_1^{k_1}\cdot\cdots\cdot x_n^{k_n})\cdot(x_1^{j_1}\cdot\cdots\cdot x_n^{j_n})\\ &\equiv_{\rho_1}(x_i^c\cdot x_i)\cdot x_1^{j_1}\cdot x_{i-1}^{j_{i-1}}\cdot x_{i+1}^{j_{i+1}}\cdot\cdots\cdot x_n^{j_n}\\ &\equiv_{\rho_1}\theta\cdot x_1^{j_1}\cdot x_{i-1}^{j_{i-1}}\cdot x_{i+1}^{j_{i+1}}\cdot\cdots\cdot x_n^{j_n}\\ &\equiv_{\rho_1}\theta. \end{aligned}$$

(3). Noting that $y_k\circ y_k\equiv_{\rho_1}(x_1^{k_1}\cdot\cdots\cdot x_n^{k_n})\circ(x_1^{k_1}\cdot\cdots\cdot x_n^{k_n})\equiv_{\rho_1}(x_1^{k_1}\circ x_1^{k_1})\cdot x_2^{k_2}\cdot\cdots\cdot x_n^{k_n}$, since $x_i\circ x_i\equiv_{\rho_1}x_i$ and $x_i^c\circ x_i^c\equiv_{\rho_1}x_i^c$, we have $y_k\circ y_k\equiv_{\rho_1}x_1^{k_1}\cdot\cdots\cdot x_n^{k_n}\equiv_{\rho_1}y_k$.

(4).

$$
\begin{aligned}
\sum_{j\in A_i}^{\circ} y_j &\equiv_{\rho_1} \sum^{\circ}(x_1^{j_1}\cdot\dots\cdot x_n^{j_n}) \\
&\equiv_{\rho_1} x_i\cdot\sum^{\circ}(x_1^{j_1}\cdot\dots\cdot x_{i-1}^{j_{i-1}}\cdot x_{i+1}^{j_{i+1}}\cdot\dots\cdot x_n^{j_n}) \\
&\equiv_{\rho_1} x_i\cdot(x_1\cdot\sum^{\circ}(x_2^{j_2}\cdot\dots\cdot x_n^{j_n})\circ\sum^{\circ}x_1^c\cdot(x_2^{j_2}\cdot\dots\cdot x_n^{j_n})) \\
&\equiv_{\rho_1} x_i\cdot[(x_1\sum^{\circ}(x_2^{j_2}\cdot\dots\cdot x_n^{j_n}))\circ(x_1^c\sum^{\circ}(x_2^{j_2}\cdot\dots\cdot x_n^{j_n}))] \\
&\equiv_{\rho_1} x_i\cdot(x_1\circ x_1^c)\cdot(\sum^{\circ}x_2^{j_2}\cdot\dots\cdot x_n^{j_n}) \\
&\equiv_{\rho_1} x_i\cdot(\sum^{\circ}x_2^{j_2}\cdot\dots\cdot x_n^{j_n}) \\
&\quad\dots \\
&\equiv_{\rho_1} x_i.
\end{aligned}
$$

(5).

$$
\begin{aligned}
\sum_{j\in A_i^c}^{\circ} y_j &\equiv_{\rho_1} \sum^{\circ}(x_1^{j_1}\cdot\dots\cdot x_n^{j_n}) \\
&\equiv_{\rho_1} x_i^c\cdot\sum^{\circ}(x_1^{j_1}\cdot\dots\cdot x_{i-1}^{j_{i-1}}\cdot x_{i+1}^{j_{i+1}}\cdot\dots\cdot x_n^{j_n}) \\
&\equiv_{\rho_1} x_i^c\cdot(x_1\cdot\sum^{\circ}(x_2^{j_2}\cdot\dots\cdot x_n^{j_n})\circ x_1^c\cdot\sum^{\circ}x_1^c(x_2^{j_2}\cdot\dots\cdot x_n^{j_n})) \\
&\equiv_{\rho_1} x_i^c\cdot[(x_1\sum^{\circ}(x_2^{j_2}\cdot\dots\cdot x_n^{j_n}))\circ(x_1^c\sum^{\circ}(x_2^{j_2}\cdot\dots\cdot x_n^{j_n}))] \\
&\equiv_{\rho_1} x_i^c\cdot(x_1\circ x_1^c)\cdot\sum^{\circ}x_2^{j_2}\cdot\dots\cdot x_n^{j_n}) \\
&\equiv_{\rho_1} x_i^c\cdot(\sum^{\circ}x_2^{j_2}\cdot\dots\cdot x_n^{j_n}) \\
&\dots \\
&\equiv_{\rho_1} x_i^c.
\end{aligned}
$$

(6).

$$
\begin{aligned}
\sum_{j\in D}^{\circ} y_j &\equiv_{\rho_1} \sum_{j\in D}^{\circ}(x_1^{j_1}\cdot\cdots\cdot x_n^{j_n}) \\
&\equiv_{\rho_1}[(x_1\cdot\sum_{j\in D}^{\circ}(x_2^{j_2}\cdot\cdots\cdot x_n^{j_n})]\circ[(x_1\cdot\sum_{j\in D}^{\circ}(x_2^{j_2}\cdot\cdots\cdot x_n^{j_n})] \\
&\equiv_{\rho_1}(x_1\circ x_1^c)\cdot(\sum_{j\in D}^{\circ} x_2^{j_2}\cdot\cdots\cdot x_n^{j_n}) \\
&\equiv_{\rho_1}(x_2\circ x_2^c)\cdot(\sum_{j\in D}^{\circ} x_3^{j_3}\cdot\cdots\cdot x_n^{j_n}) \\
&\quad\cdots \\
&\equiv_{\rho_1} x_n\circ x_n^c \\
&\equiv_{\rho_1} 1.
\end{aligned}
$$

(7). Noting that $1\circ y_k \equiv_{\rho_1} \sum_{j\in D}^{\circ} y_j\circ y_k$, since $y_k\circ y_k\equiv_{\rho_1} y_k$, we have $1\circ y_k\equiv_{\rho_1} \sum_{j\in D}^{\circ} y_j \equiv_{\rho_1} 1$.

(8). Noting that $1\circ 1\equiv_{\rho_1} 1\circ\sum_{j\in D}^{\circ} y_j$, as $1\circ y_k\equiv_{\rho_1} 1$, we can get $1\circ 1\equiv_{\rho_1} 1$.

Next, we prove $\equiv_{\rho_1}\subseteq\equiv_{\rho_2}$ . The proof will be divided into 5 steps.

(1). Noting that $x_i^c\cdot x_i \equiv_{\rho_2} \sum_{j\in A_i^c}^{\circ} y_j\cdot\sum_{j\in A_i}^{\circ} y_j$, if $y_k\cdot y_j\equiv_{\rho_2}\theta, k\neq j$, we have $x_i^c\cdot x_i\equiv_{\rho_2}\theta$.

(2). Noting that $x_i^c\circ x_i\equiv_{\rho_2}\sum_{j\in A_i^c}^{\circ} y_j\circ\sum_{j\in A_i}^{\circ} y_j$, since$A_i\cap A_i^c=\varnothing, \sum_{j\in D}^{\circ} y_j\equiv_{\rho_2} 1$, we can obtain $x_i^c\circ x_i\equiv_{\rho_2}\sum_{j\in D}^{\circ} y_j\equiv_{\rho_2} 1$.

(3). Noting that $x_i\circ x_i\equiv_{\rho_2}\sum_{j\in A_i}^{\circ} y_j\circ\sum_{j\in A_i}^{\circ} y_j$, as $y_k\circ y_k\equiv_{\rho_2} y_k$, we have $x_i\circ x_i\equiv_{\rho_2}\sum_{j\in A_i}^{\circ} y_j\equiv_{\rho_2} x_i$.

(4). The proof for $x_i^c\circ x_i^c\equiv_{\rho_1} x_i^c$ is similar to the proof of 3, and omitted.

(5). Let $A_l$ and $A_l^c$ be defined as in (3.2), (3.3). Denote $x_l^{k_l}\equiv_{\rho_2}\sum_{v\in A_l^{k_l}}^{\circ} y_v$, where $l\in\{1,2,\cdots,n\}$,

$$
A_l^k=\begin{cases} A_l & k=1 \\ A_l^c & k=0 \end{cases} \tag{3.4}
$$

Since $x_l\equiv_{\rho_2}\sum_{v\in A_l}^{\circ} y_v$, $x_l^c\equiv_{\rho_2}\sum_{v\in A_l^c}^{\circ} y_v$, we have

$$
\begin{aligned}
x_1^{k_1}\cdot x_2^{k_2}\cdot\cdots\cdot x_n^{k_n} &\equiv_{\rho_2}\sum_{v\in A_1^{k_1}}{}^{\circ} y_v\cdot\cdots\cdot\sum_{v\in A_n^{k_n}}{}^{\circ} y_v \\
&\equiv_{\rho_2}\sum_{j_1\in A_1^{k_1}}{}^{\circ} y_{j_1}\cdot\sum_{j_2\in A_2^{k_2}}{}^{\circ} y_{j_2}\cdot\cdots\cdot\sum_{j_n\in A_n^{k_n}}{}^{\circ} y_{j_n}.
\end{aligned}
$$

Since $y_k \cdot y_j (k \neq j) \equiv_{\rho_2} \theta$, if $y_{j_1} \cdot y_{j_2} \cdot \cdots \cdot y_{j_n} \neq \theta$, we have $j_1 = j_2 = \cdots = j_n$ and $j_1 = j_2 = \cdots = j_n \in \bigcap_{l=1}^{n} A_l^{k_l}$, and hence

$$\bigcap_{l=1}^{n} A_l^{k_l} = \{j\}, \quad where\ j = \sum_{l=1}^{n} k_l 2^{l-1},$$

which implies

$$x_1^{k_1} \cdot x_2^{k_2} \cdot \cdots \cdot x_n^{k_n} \equiv_{\rho_2} \sum_{j_l \in A_l^{k_l}}^{\circ} y_j \cdot y_j \cdot \cdots \cdot y_j$$
$$\equiv_{\rho_2} y_j.$$

The proof is completed. □

In order to calculate the Gröbner-Shirshov bases of $kRig[\tilde{X} \cup Y | \equiv_{\rho_2}]$, we first define a monomial ordering on $Rig[\tilde{X} \cup Y]$. Let us order $\tilde{X} \cup Y = \{x_1, x_2, \cdots, x_n, x_1^c, x_2^c, \cdots, x_n^c, y_0, \cdots, y_{m-1}\}$ generators:

$$x_1 > x_2 > \cdots > x_n > x_1^c > x_2^c > \cdots > x_n^c > y_0 > \cdots > y_{m-1}$$

then any element of $[\tilde{X} \cup Y]$ has a unique form $u = a_1 \cdot a_2 \cdot \cdots \cdot a_n$, where $a_i \in \tilde{X} \cup Y, a_1 \leqslant a_2 \leqslant \cdots \leqslant a_n n \geqslant 0$, and $u = 1$ if $n = 0$.

We order $[\tilde{X} \cup Y]$ as follows: for any $a, b \in [\tilde{X} \cup Y]$, if one of the sequences is not a prefix of order, then lexicographically; if the sequence of $a$ is a prefix of the sequence of $b$, then $a < b$.

For any $w \in Rig[\tilde{X} \cup Y]$, $w$ can be uniquely expressed as $w = u_1 \circ u_2 \circ \cdots \circ u_n$ where $u_i \in [\tilde{X} \cup Y]$. Denote $wt(w) = (u_n, u_{n-1}, \cdots, u_1)$.

We order $Rig[\tilde{X} \cup Y]$ as follows: for any $u, v \in Rig[\tilde{X} \cup Y]$, if one of the sequences is not a prefix of order, then

$$u < v \Rightarrow wt(u) < wt(v) \quad lexicographically;$$

if the sequence of $u$ is a prefix of the sequence of $v$, then $u < v$.

**Theorem 3.1.** *Let the ordering on $Rig[\tilde{X} \cup Y]$ be as above. Then $kRig[\tilde{X} \cup Y | \{(x_i \cdot x_i^c, \theta), (x_i \circ x_i^c, 1), (x_i \circ x_i, x_i), (x_i^c \circ x_i^c, x_i^c), (y_k, x_1^{k_1} \cdot \cdots \cdot x_n^{k_n})\}] = kRig[\tilde{X} \cup Y | R_1]$ and $R_1$ is a Gröbner-Shirshov basis in $kRig[\tilde{X} \cup Y]$, where $R_1$ consists of the following relations:*

$(r_1)$. $y_k \cdot y_k = y_k$,

$(r_2)$. $y_k \cdot y_j = \theta\ (k \neq j)$,

$(r_3)$. $y_k \circ y_k = y_k$,

$(r_4)$. $x_i = \sum_{j \in A_i}^{\circ} y_j\ (i \in \{1, 2, \cdots, n\})$, *where $A_i$ is defined as in (3.2),*

$(r_5)$. $x_i^c = \sum_{j \in A_i^c}^{\circ} y_j\ (i \in \{1, 2, \cdots, n\})$, *where $A_i^c$ is defined as in (3.3),*

$(r_6)$. $\sum_{j \in D}^{\circ} y_j = 1$,

$(r_7)$. $1 \circ y_k = 1$,

$(r_8)$. $1 \circ 1 = 1$.

*Proof.* We denote $\alpha \wedge \beta$ the composition of the type $(r_\alpha)$ and type $(r_\beta)$, let us check all the possible compositions.

For $1\wedge 1$, the ambiguity $w$ of all possible composition is : 1) $y_k \cdot y_k \cdot y_j \cdot y_j (k \neq j)$.

For $1 \wedge 2$, the ambiguities $w$ of all possible compositions are : 2) $y_k \cdot y_k \cdot y_j$; 3)$y_k \cdot y_k \cdot y_j \cdot y_l$ (where $k, j, l$ are not equal to each other.).

For $1\wedge 3$, the ambiguities $w$ of all possible compositions are : 4) $(y_k \cdot y_k) \circ (y_k \cdot y_k)$; 5) $(y_k \cdot y_k \cdot y_j) \circ (y_k \cdot y_k \cdot y_j)(k \neq j)$.

For $1 \wedge 4$, the ambiguity $w$ of all possible composition is : 6) $y_k \cdot y_k \cdot x_i$.

For $1 \wedge 5$, the ambiguity $w$ of all possible composition is : 7) $y_k \cdot y_k \cdot x_i^c$.

For $1 \wedge 6$, the ambiguities $w$ of all possible compositions are : 8) $\sum\limits_{j\in D}^{\circ}(y_j \cdot y_k)$; 9) $\sum\limits_{j\in D}^{\circ}(y_j \cdot y_k \cdot y_k)(k \neq j)$.

For $1 \wedge 7$, the ambiguities $w$ of all possible compositions are : 10) $(y_j \cdot y_j) \circ (y_k \cdot y_j \cdot y_j)(k \neq j)$; 11) $y_k \circ (y_k \cdot y_k)$; 12) $(y_k \cdot y_k) \circ (y_k \cdot y_k \cdot y_k)$.

For $1 \wedge 8$, the ambiguity $w$ of all possible composition is : 13) $(y_k \cdot y_k) \circ (y_k \cdot y_k)$.

For $2 \wedge 2$, the ambiguities $w$ of all possible compositions are : 14)$y_k \cdot y_j \cdot y_l$; 15) $y_k \cdot y_j \cdot y_l \cdot y_m$ (where $k, j, l$ and $m$ are not equal to each other.).

For $2\wedge 3$, the ambiguities $w$ of all possible compositions are : 16) $(y_k \cdot y_j) \circ (y_k \cdot y_j)$; 17) $(y_k \cdot y_j \cdot y_l) \circ (y_k \cdot y_j \cdot y_l)$ (where $k, j, l$ are not equal to each other.).

For $2 \wedge 4$, the ambiguity $w$ of all possible composition is : 18) $y_k \cdot y_k \cdot x_i$.

For $2 \wedge 5$, the ambiguity $w$ of all possible composition is : 19) $y_k \cdot y_k \cdot x_i^c$.

For $2 \wedge 6$, the ambiguities $w$ of all possible compositions are : 20) $\sum\limits_{j\in D}^{\circ}(y_j \cdot y_k)$; 21) $\sum\limits_{j\in D}^{\circ}(y_j \cdot y_k \cdot y_l)$ (where $k, j, l$ are not equal to each other.).

For $2 \wedge 7$, the ambiguities $w$ of all possible compositions are : 22) $y_j \circ (y_k \cdot y_j)$; 23) $(y_k \cdot y_j) \circ (y_k \cdot y_j \cdot y_l)$ $(k \neq j)$.

For $2\wedge 8$, the ambiguity $w$ of all possible composition is : 24) $(y_k \cdot y_j) \circ (y_k \cdot y_j)(k \neq j)$.

For $3 \wedge 3$, the ambiguities $w$ of all possible compositions are : 25) $(y_k \cdot y_j) \circ (y_k \cdot y_j)$,$(k \neq j)$; 26) $y_k \circ y_k$.

For $3 \wedge 4$, the ambiguity $w$ of all possible composition is : 27) $(x_i \cdot y_k) \circ (x_i \cdot y_k)$.

For $3 \wedge 5$, the ambiguity $w$ of all possible composition is : 28) $(x_i^c \cdot y_k) \circ (x_i^c \cdot y_k)$.

For $3 \wedge 6$, the ambiguities $w$ of all possible compositions are : 29) $\sum\limits_{j\in D}^{\circ} y_j \circ y_k$; 30) $\sum\limits_{j\in D}^{\circ}(y_j \cdot y_k) \circ (y_l \cdot y_k)(k \neq j, k \neq l)$.

For $3 \wedge 7$, the ambiguities $w$ of all possible compositions are :

31) $y_k \circ y_k \circ (y_k \cdot y_k)$; 32) $y_k \circ y_k \circ (y_k \cdot y_j)(k \neq j)$; 33) $y_k \circ (y_k \cdot y_j) \circ (y_k \cdot y_j)$, $(k \neq j)$; 34) $1 \circ y_k \circ y_k$.

For $3 \wedge 8$, the ambiguity $w$ of all possible composition is : 35) $y_k \circ y_k$.

For $4 \wedge 4$, the ambiguity $w$ of all possible composition is : 36) $x_i \cdot x_j$.

For $4 \wedge 5$, the ambiguity $w$ of all possible composition is : 37) $x_i \cdot x_i^c$.

For $4 \wedge 6$, the ambiguity $w$ of all possible composition is : 38) $\sum\limits_{j\in D}^{\circ}(y_j \cdot x_i)$.

For $4 \wedge 7$, the ambiguity $w$ of all possible composition is : 39) $x_i \circ (y_k \cdot x_i)$.

For $4 \wedge 8$, the ambiguity $w$ of all possible composition is : 40) $x_i \circ x_i$.

For $5 \wedge 5$, the ambiguity $w$ of all possible composition is : 41) $x_i^c \cdot x_j^c$.

For $5 \wedge 6$, the ambiguity $w$ of all possible composition is : 42) $\sum_{j\in D}^{\circ}(y_j \cdot x_i^c)$.

For $5 \wedge 7$, the ambiguity $w$ of all possible composition is : 43) $x_i^c \circ (y_j \cdot x_i^c)$.

For $5 \wedge 8$, the ambiguity $w$ of all possible composition is : 44) $x_i^c \cdot x_i^c$.

For $6 \wedge 6$, the ambiguity $w$ of all possible composition is :

45) $(y_k \cdot y_j) \circ \sum_{l\neq j}^{\circ}(y_l \cdot y_k) \circ \sum_{m\neq k}^{\circ}(y_m \cdot y_j)$.

For $6 \wedge 7$, the ambiguities $w$ of all possible compositions are : 46) $\sum_{j\in D}^{\circ}(y_k \cdot y_j) \circ y_j (k \neq j)$; 47) $\sum_{j\in D}^{\circ} y_j \circ 1$.

For $6 \wedge 8$, the ambiguity $w$ of all possible composition is : 48) $y_l \circ \sum_{j\in D}^{\circ} y_j$.

For $7 \wedge 7$, the ambiguities $w$ of all possible compositions are :

49) $1 \circ y_k \circ (y_k \cdot y_k)$; 50) $y_k \circ y_j \circ (y_k \cdot y_j)$; 51) $1 \circ y_j \circ (y_k \cdot y_j)$; 52) $1 \circ y_k \circ y_j (k \neq j)$.

For $7 \wedge 8$, the ambiguities $w$ of all possible compositions are : 53) $1 \circ 1 \circ y_k$; 54) $1 \circ y_k \circ y_k$.

For $8 \wedge 8$, the ambiguity $w$ of all possible composition is : 55) $1 \circ 1$.

Next, we have to check that all these compositions are trivial mod$(S, w)$.

Here, for example, we just check 2), 6), 11), 40), 43), 48), 53) and 54). Others can be proved similarly.

For 2), let $f = y_k \cdot y_k - y_k, g = y_k \cdot y_j - \theta$. Then $w = y_k \cdot y_k \cdot y_j$ and

$$\begin{aligned}(f, g)_w =&(y_k \cdot y_k - y_k) \cdot y_j - (y_k \cdot y_j - \theta) \cdot y_k \\ =&\theta \cdot y_k - y_k \cdot y_j \\ \equiv&0.\end{aligned}$$

From this we have the relation $r_2$.

For 6), let $f = y_k \cdot y_k - y_k, g = x_i - \sum_{j\in A_i}^{\circ} y_j$. Then $w = y_k \cdot y_k \cdot x_i$ and

$$\begin{aligned}(f, g)_w =&(y_k \circ y_k - y_k) \cdot x_i - (x_i - \sum_{j\in A_i}^{\circ} y_j) \cdot y_k \cdot y_k \\ =&(\sum_{j\in A_i}^{\circ} y_j) \cdot y_k \cdot y_k - y_k \cdot x_i \\ \equiv&0.\end{aligned}$$

From this we can get the relation $r_1$.

For 11), let $f = y_k \cdot y_k - y_k, g = 1 \circ y_k - 1$. Then $w = y_k \circ (y_k \cdot y_k)$ and

$$\begin{aligned}(f, g)_w =&(y_k \cdot y_k - y_k) \circ y_k - (1 \circ y_k - 1) \cdot y_k \\ =&y_k - y_k \circ y_k \\ \equiv&0.\end{aligned}$$

From this we can obtain the relation $r_3$.

For 40), let $f = x_i - \sum_{j\in A_i}^{\circ} y_j, g = 1 \circ 1 - 1$. Then $w = x_i \circ x_i$ and

$$\begin{aligned}(f,g)_w =&(x_i - \sum_{j\in A_i}^{\circ} y_j)\circ x_i - (1\circ 1 - 1)\cdot x_i\\ =&x_i - \sum_{j\in A_i}^{\circ} y_j \circ x_i\\ =&x_i - \sum_{j\in A_i}^{\circ} y_j \circ \sum_{j\in A_i}^{\circ} y_j\\ \equiv&0.\end{aligned}$$

From this we have the relation $r_4$.

For 43), let $f = x_i^c - \sum_{j\in A_i^c}^{\circ} y_j, g = 1\circ y_i - 1$. Then $w = x_i^c \circ (y_k \cdot x_i^c)$ and

$$\begin{aligned}(f,g)_w =&(x_i^c - \sum_{j\in A_i^c}^{\circ} y_j)\circ (y_k\cdot x_i^c) - (1\circ y_k - 1)\cdot x_i^c\\ =&x_i^c - \sum_{j\in A_i^c}^{\circ} y_j \circ (y_k\cdot x_i^c)\\ =&x_i^c - \sum_{j\in A_i^c}^{\circ} y_j \circ (y_k\cdot \sum_{j\in A_i^c}^{\circ} y_j)\\ =&\sum_{j\in A_i^c}^{\circ} y_j - \sum_{j\in A_i^c}^{\circ} y_j \circ (y_k\cdot \sum_{j\in A_i^c}^{\circ} y_j)\\ \equiv&0.\end{aligned}$$

From this we can obtain the relation $r_5$.

For 48), let $f = \sum_{j\in D}^{\circ} y_j - 1, g = 1\circ 1 - 1$. Then $w = y_k \circ \sum_{j\in D}^{\circ} y_j$ and

$$\begin{aligned}(f,g)_w =&(\sum_{j\in D}^{\circ} y_j - 1)\circ y_k - (1\circ 1 - 1)\cdot y_k \circ \sum_{j\neq k}^{\circ} y_j\\ =&y_k \circ \sum_{j\neq k}^{\circ} y_j - 1\circ y_k\\ =&\sum_{j\in D}^{\circ} y_j - 1\circ y_k\\ \equiv&0.\end{aligned}$$

From this we have the relation $r_6$.

For 53), let $f = 1\circ y_k - 1, g = 1\circ 1 - 1$. Then $w = 1\circ 1\circ y_k$ and

$$\begin{aligned}(f,g)_w =&(1\circ y_k - 1)\circ 1 - (1\circ 1 - 1)\circ y_k\\ =&1\circ y_k - 1\circ 1\\ \equiv&0.\end{aligned}$$

From this we can get the relation $r_8$.

For 54), let $f = 1\circ y_k - 1, g = 1\circ 1 - 1$. Then $w = 1\circ y_k \circ y_k$ and

$$
\begin{aligned}
(f,g)_w =&(1\circ y_k-1)\circ y_k-(1\circ 1-1)\cdot y_k\circ 1\\
=&y_k\circ 1-1\circ y_k\\
\equiv&0.
\end{aligned}
$$

From this we have the $r_7$.

□

**Remark 3.1.** Since $y_k\circ y_k$ is divisible by $1\circ 1$. So, the reduced Grӧbner-Shirshov basis denoted by $R_1^{comp}$ consists of the following relations:

$(r_1)$. $y_k\cdot y_k=y_k$,

$(r_2)$. $y_k\cdot y_j=\theta$ $(k\neq j)$,

$(r_3)$. $x_i=\sum^{\circ}_{j\in A_i} y_j$, $(i\in\{1,2,\cdots,n\})$, where $A_i$ is defined as in (3.2),

$(r_4)$. $x_i^c=\sum^{\circ}_{j\in A_i^c} y_j$, $(i\in\{1,2,\cdots,n\})$, where $A_i^c$ is defined as in (3.3),

$(r_5)$. $\sum^{\circ}_{j\in D} y_j=1$,

$(r_6)$. $1\circ y_k=1$,

$(r_7)$. $1\circ 1=1$.

**Remark 3.2.** In the following, we will denote $Rig[\tilde{X}]/\equiv_\rho$ by $S_c[X]$.

From Theorem 3.1 and Theorem 2.1, we have the following corollaries.

**Corollary 3.1.** *A normal form of the semiring $S_c[X]$ is the set*

$$\{1,\sum_{k\in D}^{\circ} y_k\},$$

*where $D\subsetneq\{0,1,\cdots,m-1\}$.*

**Corollary 3.2.** *For any $s\in S_c[X]$, there exists a unique $t$ such that:*

*(1) $s\cdot t=\theta$,*

*(2) $s\circ t=1$,*

*(3) $s\cdot s=s$,*

*(4) $s\circ s=s$.*

*Proof.* For any s $\in S_c[X]$, if $s=1$, then $t=\theta$; if $s\neq 1$, then we let $s=\sum^{\circ}_{i\in A} y_i$, where $A\subsetneq\{0,1,\cdots,m-1\}$, and let $t=\sum^{\circ}_{i\in\overline{A}} y_i$, where $\overline{A}=\{0,1,\cdots,m-1\}-A$. Then

$$s\cdot t=\theta,\quad s\circ t=1.$$

If

$$s\cdot(s\circ t)=(s\cdot s)\circ(s\cdot t)=s,$$

we have $s\cdot s=s$.

If

$$s\circ(s\cdot t)=(s\circ s)\cdot(s\circ t)=s\circ\theta,$$

we have $s\circ s=s$.

Next, we prove the uniqueness of $t$. Let $t_1 \neq t_2$ such that

$$s \cdot t_1 = \theta, \quad s \cdot t_2 = \theta, \quad s \circ t_1 = 1, \quad s \circ t_2 = 1,$$

then we have

$$t_1 = (s \circ t_2) \cdot t_1 = (s \cdot t_1) \circ (t_2 \cdot t_1) = t_2 \cdot t_1,$$

and

$$t_2 = (s \circ t_1) \cdot t_2 = (s \cdot t_2) \circ (t_1 \cdot t_2) = t_1 \cdot t_2,$$

These leads to a contradiction. □

**Definition 3.1.** Let $S$ be a comutative semiring with operations " $\circ$ " and multiplication " $\cdot$ ". If for any $s \in S$, there exists $t \in S$ such that the following conditions are satisfied:

(1) $s \cdot t = \theta$.
(2) $s \circ t = 1$.
(3) $s \cdot s = s$.
(4) $s \circ s = s$.

Then we call $S$ is an idempotent complement semiring, where $t$ is the complement of $s$, and denote $t = s^c$.

**Remark 3.3.** In Definition 3.1, $t$ is unique, the proof is similar to Corollary 3.2.

**Remark 3.4.** From the normal form of $S_c[X]$, it is easy to prove that $S_c[X]$ is an idempotent complement semiring.

**Remark 3.5.** $S_c[X]$ is a free idempotent complement semiring generated by $X$. We denote $\eta$: $X \to S_c[X]$, via $x_i \mapsto \bar{x_i}$, by C-D lemma and Theorem 3.1, it is easy to prove that $\eta$ is an injective. For any idempotent complement semiring $S$, for any $f : X \to S$, via $x_i \mapsto f(x_i)$, there exists a unique semiring homomorphism $\bar{f}$ such that the following graph is commutative.

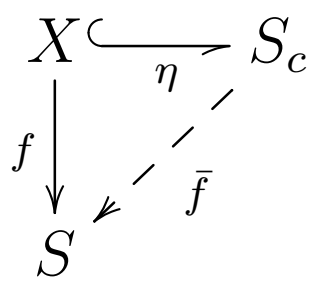

## 4. Markov semiring

Now, let $X = \{x_1, x_2, \cdots, x_n\}$, where $x_i$ is a discrete random variable, $i \in \{1, 2, \cdots, n\}$. $S_c[X]$ is the free idempotent complement semiring generated by $X$. In $S_c[X]$, we can define $f(x_{i_1} \circ x_{i_2} \circ \cdots \circ x_{i_k}) = H(x_{i_1}, x_{i_2}, \cdots, x_{i_k})$, $f(\theta) = 0$, where $H(x_{i_1}, x_{i_2}, \cdots, x_{i_k})$ is the joint entropy of $x_{i_1}, x_{i_2}, \cdots, x_{i_k}$. According to [13, Theorem 3.6], if $f$ satisfies:

$$f(x \cdot y) = f(x) + f(y) - f(x \circ y),$$

then $f$ can be uniquely extended to a real-valued function on $S_c[X]$, which is still denoted by $f$.

In fact, at this point:[12]

$$f(x_i \cdot x_j^c) = H(x_i | x_j),$$

$$f(x_i \circ x_j) = H(x_i, x_j),$$

$$f(x_i \cdot x_j^c \cdot x_k) = I(x_i; x_k | x_j),$$

where $I(x_i; x_k|x_j)$ is the mutual information between $x_i$ and $x_k$ conditioning on $x_j$. If $x_1 \to x_2 \to \cdots \to x_n$ forms a Markov chain, it equivalent to $\{f(x_1 \cdot x_2^c \cdot x_3) = I(x_1; x_3|x_2) = 0,\ f((x_1 \circ x_2) \cdot x_3^c \cdot x_4) = I((x_1, x_2); x_4|x_3) = 0, \cdots, f((x_1 \circ x_2 \circ \cdots \circ x_{n-2}) \cdot x_{n-1}^c \cdot x_n) = I((x_1, x_2, \cdots, x_{n-2}); x_n|x_{n-1}) = 0\}$. Since $I$ is the information measure, it means that $\{x_1 \cdot x_2^c \cdot x_3 = \theta,\ (x_1 \circ x_2) \cdot x_3^c \cdot x_4 = \theta, \cdots, (x_1 \circ x_2 \circ \cdots \circ x_{n-2}) \cdot x_{n-1}^c \cdot x_n = \theta\}$.

Therefore, we can consider the algebratic structure and normal form of the semiring defined as following.

**Definition 4.1.** Suppose that $X = \{x_1, x_2, \cdots, x_n\}$. $S_c[X]$ is the free idempotent complement semiring generated by $X$. If in $S_c[X]$: $x_1 \cdot x_2^c \cdot x_3 = \theta, (x_1 \circ x_2) \cdot x_3^c \cdot x_4 = \theta, \cdots, (x_1 \circ x_2 \circ \cdots \circ x_{n-2}) \cdot x_{n-1}^c \cdot x_n = \theta$, then we call $x_1 \to x_2 \to x_3 \to \cdots \to x_n$ is a Markov chain. Let $\equiv_{\rho_n}$ be the congruence relation in $S_c[X]$ generated by $\{(x_1 \cdot x_2^c \cdot x_3, \theta), ((x_1 \circ x_2) \cdot x_3^c \cdot x_4, \theta), \cdots, ((x_1 \circ x_2 \circ \cdots \circ x_{n-2}) \cdot x_{n-1}^c \cdot x_n, \theta)\}$, then we call $S_c[X]/\equiv_{\rho_n}$ a Markov semiring, denote it $M_n[X]$.

In this section, we will study algebraic expressions and their normal forms of $M_n[X]$. Now, we use Shirshov algorithm to find the Gröbner-Shirshov bases and their normal forms.

When $n = 5$, then $M_5[X]$ is generated by the following relations $R_5$:

$(r_1)$. $y_k \cdot y_k = y_k$ $(k = 1, 2, \cdots, 31)$,

$(r_2)$. $y_k \cdot y_j = \theta$ $(k \neq j, j = 1, 2, \cdots, 31)$,

$(r_3)$. $x_i = \sum^{\circ}_{j \in A_i} y_j$ $(i = 1, 2, 3, 4, 5)$, where $A_i$ is defined as in (3.2),

$(r_4)$. $x_i^c = \sum^{\circ}_{j \in A_i^c} y_j$, where $A_i^c$ is defined as in (3.3),

$(r_5)$. $\sum^{\circ}_{j \in D} y_j = 1$,

$(r_6)$. $1 \circ y_k = 1$,

$(r_7)$. $1 \circ 1 = 1$,

$(r_8)$. $x_1 \cdot x_2^c \cdot x_3 = \theta$,

$(r_9)$. $(x_1 \circ x_2) \cdot x_3^c \cdot x_4 = \theta$,

$(r_{10})$. $(x_1 \circ x_2 \circ x_3) \cdot x_4^c \cdot x_5 = \theta$.

By Shirshov algorithm, we need add to $R_5$ all nontrivial compositions, for example, for $1 \wedge 8$, the ambiguities $w$ of all possible compositions are: $w = x_1 \cdot x_2^c \cdot x_3 \cdot y_k \cdot y_k, (k = 0, \cdots, 31)$

Let $f = y_k \cdot y_k - y_k, g = x_1 \cdot x_2^c \cdot x_3 - \theta$ and

$$\begin{aligned}(f, g)_w &= (y_k \cdot y_k - y_k) \cdot (x_1 \cdot x_2^c \cdot x_3) - (x_1 \cdot x_2^c \cdot x_3 - \theta) \cdot (y_k \cdot y_k) \\ &= \theta \cdot (y_k \cdot y_k) - y_k \cdot (x_1 \cdot x_2^c \cdot x_3) \\ &= \theta - y_k \cdot (x_1 \cdot x_2^c \cdot x_3).\end{aligned}$$

Then we have:

If $k \notin A_1 \cap A_2^c \cap A_3, (f, g)_w = \theta - \theta \equiv 0$.

If $k \in A_1 \cap A_2^c \cap A_3$, for $A_1 \cap A_2^c \cap A_3 = \{5, 13, 21, 29\}$, we can get:

When $k=5$,

(11) $(f,g)_w=\theta-y_5\cdot(x_1\cdot x_2^c\cdot x_3)=\theta-y_5\cdot(y_5\circ y_{13}\circ y_{21}\circ y_{29})=\theta-y_5$.

So, we have to add the relation $(y_5,\theta)$ to $R_5$ .

Similarly, if $k\in\{13,21,29\}$, we have to add the relations $(y_{13},\theta)$, $(y_{21},\theta)$, $(y_{29},\theta)$ to $R_5$.

Therefore, we should add $y_k-\theta$ to $R_5$, where $k\in\{5,13,21,29\}$.

Likewise, we computer the compositions for $2\wedge 8,\cdots,7\wedge 8, i\wedge 9$ and $i\wedge 10$, where $i\in\{1,\cdots,10\}$, and add them to $R_5$ denote by $(r_{11})$, as follows:

$(r_1)$. $y_k\cdot y_k=y_k$,

$(r_2)$. $y_k\cdot y_j=\theta\ (i\neq j)$,

$(r_3)$. $x_i=\sum^{\circ}_{j\in A_i} y_j\ (i=1,2,3,4,5)$, where $A_i$ is defined as in (3.2),

$(r_4)$. $x_i^c=\sum^{\circ}_{j\in A_i^c} y_j$, where $A_i^c$ is defined as in (3.3),

$(r_5)$. $\sum^{\circ}_{j\in D} y_j=1$,

$(r_6)$. $1\circ y_k=1$,

$(r_7)$. $1\circ 1=1$,

$(r_8)$. $x_1\cdot x_2^c\cdot x_3=\theta$,

$(r_9)$. $(x_1\circ x_2)\cdot x_3^c\cdot x_4=\theta$,

$(r_{10})$. $(x_1\circ x_2\cdot x_3)\cdot x_4^c\cdot x_5=\theta$,

$(r_{11})$. $y_p=\theta, p\in K_5$, where $K_5=\{5,13,21,29,9,10,11,24,25,26,17,18,19,20,22,23\}$.

Furthermore, according to Shirshov algorithm, we computer nontrivial compositions of $i\wedge 11$, and repeat the procedure above, we can get the Gröbner-Shirshov basis $R_5^{comp}$, where $R_5^{comp}$ consists of the following relations:

$(r_1)$. $y_k\cdot y_k=y_k$,

$(r_2)$. $y_k\cdot y_j=\theta\ (i\neq j)$,

$(r_3)$. $x_i=\sum^{\circ}_{j\in A_i} y_j\ (i=1,2,3,4,5)$, where $A_i$ is defined as in (3.2),

$(r_4)$. $x_i^c=\sum^{\circ}_{j\in A_i^c} y_j$, where $A_i^c$ is defined as in (3.3),

$(r_5)$. $\sum^{\circ}_{j\in D-K_5} y_j=1$, where $K_5=\{5,13,21,29,9,10,11,24,25,26,17,18,19,20,22,23\}$,

$(r_6)$. $1\circ y_k=1$,

$(r_7)$. $1\circ 1=1$,

$(r_8)$. $x_1\cdot x_2^c\cdot x_3=\theta$,

$(r_9)$. $(x_1\circ x_2)\cdot x_3^c\cdot x_4=\theta$,

$(r_{10})$. $(x_1\circ x_2\cdot x_3)\cdot x_4^c\cdot x_5=\theta$,

$(r_{11})$. $y_p=\theta$, $p\in K_5$.

Since the leading terms of $(r_8),(r_9)$ and $(r_{10})$ are divisible by the leading terms of $(r_3)$, $(r_4)$. So, similar to Remark 3.1, we can get:

**Theorem 4.1.** *For $n=5$, let the monomial ordering on $Rig[\tilde{X}\cup Y]$ be as in Theorem 3.1. Then $kM_5[X]$ has a Gröbner-Shirshov basis consists of the following relations:*

$(r_1)$. $y_k\cdot y_k=y_k$,

$(r_2)$. $y_k\cdot y_j=\theta\ (k\neq j)$,

$(r_3)$. $x_i=\sum^{\circ}_{j\in A_i} y_j\ (i=1,2,3,4,5)$, *where $A_i$ is defined as in (3.2),*

$(r_4)$. $x_i^c = \sum^{\circ}_{j\in A_i^c} y_j$, *where* $A_i^c$ *is defined as in (3.3),*
$(r_5)$. $\sum^{\circ}_{j\in D-K_5} y_j = 1$, *where* $K_5 = \{5, 13, 21, 29, 9, 10, 11, 24, 25, 26, 17, 18, 19, 20, 22, 23\}$,
$(r_6)$. $1 \circ y_k = 1$,
$(r_7)$. $1 \circ 1 = 1$,
$(r_8)$. $y_p = \theta$, $p \in K_5$.

From Theorem 2.1 and Theorem 4.1, we have the following corollary.

**Corollary 4.1.** *A normal form of* $M_5[X]$ *is the set*

$$\{1, \sum^{\circ}_{k\in B} y_k\},$$

*where* $B \subsetneq \{0, 1, 2, 3, 4, 6, 7, 8, 12, 14, 15, 16, 24, 28, 30, 31\}$.

Therefore, we can obtain the algebraic structure of the Markov semiring $M_n[X]$ and its normal form when the number of random variables is $n(n \geqslant 3)$.

**Theorem 4.2.** *Let the monomial ordering on* $Rig[\tilde{X} \cup Y]$ *be as in Theorem 4.1. Then* $kM_n[X]$ *has a Gröbner-Shirshov basis consists of the following relations:*
$(r_1)$. $y_k \cdot y_k = y_k$,
$(r_2)$. $y_k \cdot y_j = \theta$ $(k \neq j)$,
$(r_3)$. $x_i = \sum^{\circ}_{j\in A_i} y_j$, *where* $A_i$ *is defined as in (3.2),*
$(r_4)$. $x_i^c = \sum^{\circ}_{j\in A_i^c} y_j$, *where* $A_i^c$ *is defined as in (3.3),*
$(r_5)$. $\sum^{\circ}_{j\in D-K_n} y_j = 1$, *where* $K_n = \bigcup_{i=1}^{n-2} ((A_1 \cup A_2 \cup \cdots \cup A_i) \cap A_{i+1}^c \cap A_{i+2})$,
$(r_6)$. $1 \circ y_k = 1$,
$(r_7)$. $1 \circ 1 = 1$,
$(r_8)$. $y_p = \theta$, $p \in K_n$.

**Corollary 4.2.** *A normal form of* $M_n[X]$ *is the set*

$$\{1, \sum^{\circ}_{k\in B} y_k\},$$

*where* $B \subseteq \{\{0, 1, \cdots, 2^n - 1\} - K_n\}$.

**Remark 4.1.** For $n = 3$ and $n = 4$ Markov semirings, results obtained from Corollary 4.2 are similar to that of [13].

**Remark 4.2.** If $x_1 \to x_2 \to x_3 \to \cdots \to x_n$ is a Markov chain, since $K_n = \bigcup_{i=1}^{n-2} ((A_n \cup A_{n-1} \cup \cdots \cup A_{n-i+1}) \cap A_{n-i}^c \cap A_{n-i-1})$, by Theorem 4.2, we can get $x_n \to x_{n-1} \to x_{n-2} \to \cdots \to x_1$ is a Markov chain.

## Acknowledgment

This work is supported by the National Social Science Fundation of China (No. 21BJY213), and also by NSF of Anhui University, China (No. KJ2021A0386, KJ2021A1034)